\documentclass[compsoc, conference, a4paper, 10pt, times]{IEEEtran}
\usepackage{filecontents}
\usepackage{comment}
\IEEEoverridecommandlockouts
\usepackage{cite}
\usepackage{amsmath,amssymb,amsfonts}
\usepackage{algorithmic}
\usepackage{graphicx}
\usepackage{booktabs}

\PassOptionsToPackage{hyphens}{url}
\usepackage[hidelinks]{hyperref}
\usepackage{textcomp}
\usepackage[table,x11names, dvipsnames]{xcolor}
\def\BibTeX{{\rm B\kern-.05em{\sc i\kern-.025em b}\kern-.08em
    T\kern-.1667em\lower.7ex\hbox{E}\kern-.125emX}}

\newif\ifstatus
\statustrue % Comments on
\newcommand{\todolater}[1]{}

\begin{document}

\title{Understanding Crypter-as-a-Service in a popular underground marketplace}

\author{\IEEEauthorblockN{Alejandro de la Cruz Alvarado}
\IEEEauthorblockA{\textit{Universidad Carlos III de Madrid}\\
Madrid, Spain \\
alcruza@pa.uc3m.es}
\and
\IEEEauthorblockN{Sergio Pastrana Portillo}
\IEEEauthorblockA{\textit{Universidad Carlos III de Madrid}\\
Madrid, Spain \\
spastran@inf.uc3m.es}
}

\maketitle

\begin{abstract}
    
        \textit{Crypters} are pieces of software whose main goal is to transform a target binary so it can avoid detection from Anti Viruses (AVs from now on) applications. They work similar to packers, by taking a malware binary $M$ and applying a series of modifications, obfuscations and encryptions to output a binary $M'$ that evades one or more AVs. The goal is to remain fully undetected, or FUD in the hacking jargon, while maintaining its (often malicious) functionality. In line to the growth of commoditization in cybercrime, the Crypter-as-a-Service model (CaaS) has gained popularity, in response to the increased sophistication of detection mechanisms. In this business model, customers receive an initial crypter which is soon updated once becomes detected by anti-viruses. 

        This paper provides the first study on an online underground market dedicated to Crypter-as-a-Service.
        We compare the most relevant products in sale, analyzing the existent social network on the platform and comparing the different features that they provide. We also conduct an experiment as a case study, to validate the usage of one of the most popular crypters sold in the marketplace, and compare the results before and after crypting binaries (both benign software and malware), to show its effectiveness when evading antivirus engines.
\end{abstract}


\section{Introduction}

    In the recent years, there has been a large increase in the sophistication of cyber defense systems, with multiple detection improvements, involving artificial intelligence and crowdsourced projects where professionals share knowledge for a common benefit. The major example in the realm of malware detection is the cooperative-aimed project VirusTotal,\footnote{\url{https://www.virustotal.com}} where anyone can upload a binary to gain insights from multiple AV engines, and in turn it offers open source intelligence (OSINT) from the results. These advances motivate cyber criminals to evolve as well. In the case of malware, developers have adapted to provide stealth infection by means of binary obfuscation. Concretely, this approach focuses on hindering the reversing and static analysis of a piece of software, so it remains undetected at the eyes of antivirus. Accordingly, due to a widely use of binary obfuscation, malware analysts require to understand and reverse the obfuscation process before gaining access to the actual binary code~\cite{obfuscation}.
    
    Binary obfuscation is often conducted by \textit{packers}. These are not particular for malware, and are also widely used in the commercial space. For example, when a product is going to be publicly available such as videogames, they typically combine obfuscation with some sort of encryption, semantic modifications and redundancies to hide knowledge from reverse engineers.
    Packers can be customized, or even made ad hoc by the developer for each binary, but in most cases, it is a standard product sold in some marketplace. Crypters, our focus in this study, is a type of packer that cryptographically modifies a binary to evade antivirus engines. Due to high technical skills required to build a custom crypter, and provided that these are mostly used for malicious purposes, crypters are usually traded in underground marketplaces and forums~\cite{crimeBB}~\cite{plugPlay}.
    
    Cybercrime is a growing issue, with an increased prevalence since the 2020 pandemic~\cite{buil2021cybercrime}. Indeed, the availability of low-level hacking tools in public underground forums lowers the barrier for non-expert users (or script-kiddies as they are known in the community), have ease the access to exploits and hacking material that can be used for malicious intentions~\cite{crimeBB,plugPlay}. It is thus important to study how it evolves and what is the current status of the different activities~\cite{hughes2024art}.
    
    Despite the high amount of academic research on cybercrime, with some works specializing in technical details of  obfuscation tools~\cite{avEvasion,vsembera2021cybercrime,muralidharan2022file}, there is a gap of studies focused on the Crypter-as-a-Service (CaaS) ecosystem, from the marketing and operation perspective.
    
    To fill this gap, in this paper we analyze the marketplace of crypters traded in a popular english-speaking underground forum with a specific subforum dedicated to crypters, i.e., HackForums. We analyze the most common properties, techniques used to advertise their products and attract new customers, and the differences between them. We also analyze the social network formed by the actor involved in creating, buying and interacting with the multiple products available. Finally, we conduct an analysis with one of the top crypters sold, showing how it bypasses antiviruses by comparing a ``crypted'' binary with the original, non-obfuscated binary. To this end, we have first developed a custom crawler for HackForums marketplace to store the information timestamp of creation and last update, content, users, etc. 
    
    Overall, the main contributions of this paper are the following:
    \begin{itemize}
        \item We provide the first quantitative analysis of the CaaS ecosystem in a popular underground forum with a long lasting crypter-focused subforum along with a deeper study of the products sold and the activity in the marketplace.
        \item We describe our crawler and data collection method, that allows us to obtain a total of 1,492 threads or posts and 128,384 comments in those posts along with the information of 17,751 users.
        \item We conduct an analysis of the collected data, including a social network analysis of the marketplace to find market niches, the most relevant users and differences between products being sold.
        \item Finally, as a case study, we empirically validate the usage of one of the top crypter products sold, to compare the AV results of a binary before and after running the crypter.
    \end{itemize}

    Finally, to foster research in the area and allow for reproducibility, we open-source the crawler and the analysis scripts in our anonymized repository~\cite{repo}.

    The rest of the paper is structured as follows. First, Section~\ref{sec:crypter-as-a-service} provides an overview of the crypter-as-a-service model and how it is operated in the underground economy. Then, Section~\ref{sec:crawler} describes the data collection process by means of a custom crawler for HackForums, bypassing its protections against bot activity and the data structure used to store the data. In Section~\ref{sec:analysis} we analyze the collected data and discuss the main insights gathered from the results. Then, Section~\ref{sec:technical} shows a case study through the usage of one of the top crypters being sold in the platform and a technical analysis of a crypted binary. Finally, we discuss related work in Section~\ref{sec:related-work}, and describe the main conclusions in Section~\ref{sec:conclusions}.

\section{The Crypter-as-a-service model}
\label{sec:crypter-as-a-service}

    In the recent years, the cybercrime landscape has evolved considerably~\cite{crimeBB}. To a great extent, this is due to the commoditization of products and services available in online market and forums~\cite{plugPlay}. In this scenario, several criminal companies have grown, whose main goal is typically to gain some economic benefit. This foster the rise of Business-to-Business (B2B) services, where criminals provide the necessary services and products to others, cultivating criminal `entrepreneurship'~\cite{bohme2021silicon}. 
    In the case of malware infection, it requires at least 3 main components or services. First, the malware development (e.g., acquiring a customized crytomining malware~\cite{pastrana2019first} or banking trojans~\cite{paquet2022motivations}). Second, the malware spreading (e.g., buying `installs' from botnet operators to spread the malware across several victims~\cite{caballero2011measuring}). And third, the malware obfuscation, so it remains undetected and allows for persistence on the infected computers. The latter, which is the focus of this paper, is often conducted by means of crypters.

    A crypter is commonly composed by two parts: the ``builder'' and the ``stub''. The former is responsible for the encryption and obfuscation of the binary and some personalized tuning. Since crypters are commercial products, potentially targeted for users with little or no technical knowledge, the builder usually includes a graphical user interface (GUI), with step-by-step instructions, and even an allegedly appropriate customer service. Other complementary functionalities advertised are icon and name personalization, delayed start and persistence (see Section~\ref{sec:analysis}). The ``stub'', which is the most important component of a crypter, is the file being generated as output and in charge of the decryption and execution of the original malware. Since this is the piece that will be installed on a victims' computer, it requires to evade AVs~\cite{crypters}.
    
    The builder disguises the malware by obfuscating it using different means, most commonly by shuffling instructions, encrypting it completely and hiding them in a benign (undetected) file or ``stub''. Then this stub decrypts the actual binary in memory at runtime, and executes the original functionality with a technique known as Dynamic Forking~\cite{stubs}, which basically invokes a suspended process and then gets replaced by the malicious process to be executed. Since the stub is a static part, it can be detected by antivirus. Indeed, as soon as a stub is fingerprinted by any AV, any malware using it is quickly detected and the crypter becomes useless. Accordingly, most of the crypters' providers the stub is updated when needed, to guarantee that it remains `fully undetected'' (or FUD, in the underground jargon). This is the reason for the emergence of the `crypter-as-a-service' business model.

\begin{figure}[h!]
      \centering
      \includegraphics[width=\linewidth]{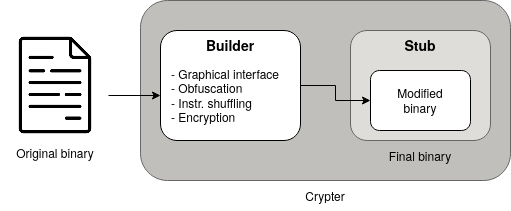}
      \caption{Components of a crypter}
      \label{fig:crypterStages}
    \end{figure}

    Figure~\ref{fig:crypterStages} depicts the different stages a crypter goes through from the original binary to a shuffled and obfuscated set of instructions split inside the stub or ``FUD'' binary.

    Crypters can be classified into scantime crypter or runtime crypter, based on the stage when the binary needs to be hidden. A scantime crypter remains static in disk, thus bypassing traditional antiviruses, while a runtime crypter is loaded in memory and needs to be more complex to evade other tools and defenses such as Host IDS, Windows AMSI or Endpoint Detection and Response (EDR) systems.

\begin{figure}[ht]
      \centering
      \includegraphics[width=\columnwidth]{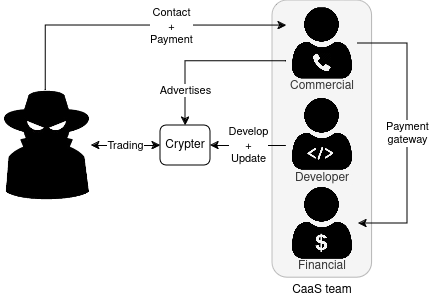}
      \caption{B2B relations in the crypting-as-a-service model}
      \label{fig:b2b}
    \end{figure}

    The crypter-as-a-service paradigm is based on other crime-as-a-service models~\cite{MANKY20139} with several roles which might be operated by the same individual, or they can be diversified with a entrepreneurial approach~\cite{bohme2021silicon} (see Figure~\ref{fig:b2b}). This model is reinforced by our quantitative analysis, detecting multiple duplicated products and authors of posts claiming not to be the developer of the crypter but the reseller. We describe the three most common roles involved:
    \begin{enumerate}
        \item The \textit{crypter developer} team is in charge of programming the main technical components of the crypter. It also updates the builder when the stub becomes detected by the AV industry.
        \item The \textit{commercial} team advertises the service in underground forums. It can be also responsible for the development and maintenance of the main web portal where the crypter is being sold (though this can be externalize to another \textit{web developer}). Its main goal is to attract customers, and also to provide customer service when needed. It demands the payments to the customers, which are in turn managed by the finance team as discussed next.
        \item The \textit{finance} team creates and maintains the payment platform. This is often based on cryptocurrencies (e.g., Bitcoin, Ethereum or Monero), or by other online payment systems popular in criminal businesses, such as PayPal or Amazon Gift Cards. It must keep the payments anonymous, and also is responsible for the money laundering.
    \end{enumerate}

    The never ending race between antivirus and malware obfuscation makes the crypter-a-service business a relevant aspect in the cyber-criminal landscape, since this is a key component to maintain the kill-chain during malware infection. Accordingly, there are underground marketplaces and forums specialized in the trading of these services. One of the largest belongs to HackForums, which is the focus of our study as we detail in the following sections.

\section{Data collection}
\label{sec:crawler}
    
    This study focuses on the marketplace of crypters traded in HackForums. Thus, the first step is to crawl and scrape all the existent threads in the forum and dump the information into a database for offline analysis over the collected data.
    
    \subsection{Data structure}

        HackForums is a public forum on the clear web discussing cybersecurity, with numerous debates and tutorials on several topics like hacking, botnets, etc. However, it has also been the home for various cybercriminals ~\cite{pastrana2018characterizing}. We focus on the marketplace under the ``Cryptography and Encryption'' section to get all the threads announcing crypters. Marketplaces in HackForums are structured in different threads where the original post is the announcement of a product, in this case, a crypter, and in that same thread users can leave comments with questions regarding the product, request free trials, test the results against AVs or rate the product. The product announcement might be formated as a pamphlet with the main characteristics, the prize and in general any information that might be useful to sell the product to customers. On top of that, the thread creator might also add a textual description of the product, share any social media profile where users can reach to purchase the crypter or any other information.

        The threads are listed without the need for an account, and even though the information is very limited, one can get the names of the posts, the name of the creator, the number of views and when was the last interaction. This provides enough information to get a grasp of the market offer. But to get the whole information, we need to access the posts (replies) of the thread. This poses an additional challenge, since accesing the contents of a thread requires to be logged into the website, and with the amount of captchas and bot detections it is hard to automate~\cite{cyberCrime}. In next section we describe our method to bypass the access barriers. 
        Once we can access each post, we extract the following information: the original post content, the image (pamphlet) used to announce the crypter, if existent, and all the posts of the thread, left by the users as replies. These interactions between users and posts will let us visualize the social network of crypter producers and consumers.
        
        Overall, we collected 1 492 posts from 382 different users and 128 384 comments in those posts from a total of 17 751 users. We conducted our crawling on April 2023, collecting historical data that spans for 13 years.

    \subsection{Crawling an antibot forum}
        Hackforums implements various antibot mechanisms. During our research, some of our requests were rejected, we got some accounts locked and even some IPs blocked from the website. In order to solve all these inconveniences and reduce the risk of being blocked, we implement the countermeasures discussed next.
        
        Due to the high volume of requests made by crawlers, we configure \textit{Scrapy}, a Python library used for our scraper, to work with \textit{Tor} and \textit{privoxy} to rotate IPs each time we get a bad response from the server, thus solving the IP blockage issue.
        
        The forum also detects artificial activity and out-of-the-ordinary behaviors, but using the API from ScrapeOps we managed to modify the default User Agent, the time between requests and use its proxy to balance between multiple endpoints making our crawler seem like trustworthy activity.
        
        To run the scraper using a valid session, we first open a browser with \textit{playwright}, manually log in, and then run the crawler to automatically fetch the required pages. Also, to minimize the requests duing the handling of potential errors, the crawler downloads a local copy of all the html webpage collected. This way, when an error occurs in the scraping of a particular site, we could re-run it offline, avoiding repeating the process which otherwise would risk the account to be blocked, and also making the process overall faster.

\section{Data Analysis}
\label{sec:analysis}
    
    We first analyze the collected data to characterize the information from the announced data itself, providing a generic measurement (e.g., the frequency of post creations to the most frequently used languages or words). We then analyse the most popular crypters (measured by number of views), to have a more precise conclusion over the most popular ones. Finally, we study the social network behind the forum.

    The goal of these analyses is to better understand to activity in the subforum, social network beneath it and highlight common properties and differences between crypters advertised in HackForums to find tendencies and specializations in the marketplace.

    \subsection{Global measurement}

        The first approach analyzes the languages used in both the comments and posts in case there were region-specific crypters or maybe types of crypters more suited for some conditions than others.
        To do this we used \textit{polyglot}, a popular natural language Python library that supports massive multilingual analysis, and which is commonly used in the field of data science. The forum is mainly popular across the US, Europe and other english-speaking countries such as India. Accordingly, to reach as many customers as possible almost the total amount of posts (99.9\%) and comments (99.99\%) were written in English.

        We next analyze the most frequent words, which help to derive interesting insights on the offered features. We first took the number of occurrences for every word found in the threads of the marketplace, removed the non-significant words for this study such as pronouns, articles, etc. and limited our study to words with over one thousand appearances.
    
\begin{figure}[h]
          \centering
          \includegraphics[width=\linewidth]{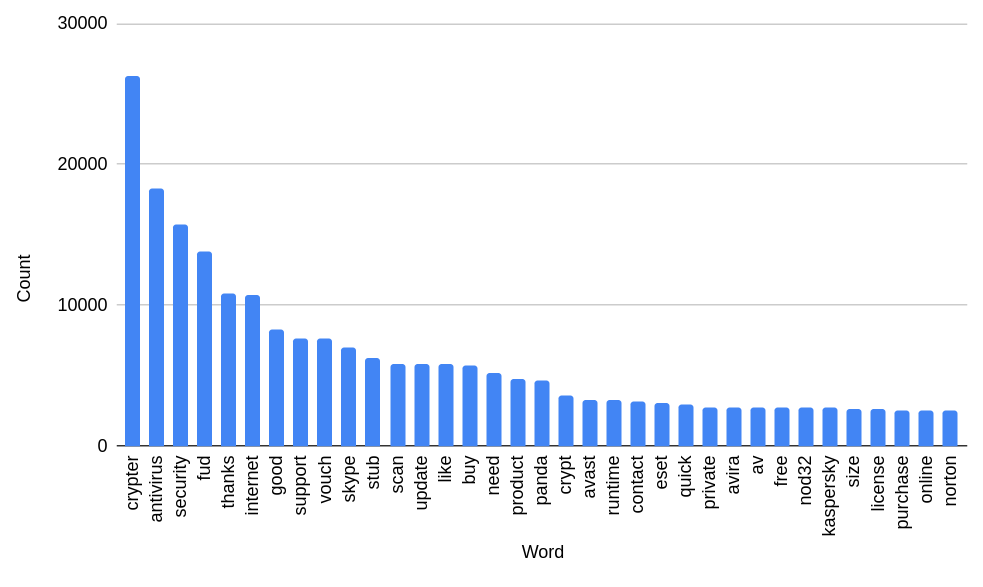}
          \caption{Word histogram}
          \label{fig:words}
        \end{figure}

        As seen in Figure~\ref{fig:words}, the resulting histogram, some of the most recurrent words are related to the products sold (\textit{``crypter'', ``FUD''}), ratings of the product, users asking for free \textit{``vouches''}, \textit{``trials''} or \textit{``licenses''}, the type of stub used (\textit{``size''}, \textit{``private''}, \textit{``runtime''}), topics related to antiviruses (\textit{``antivirus''} being the second most used word, but also \textit{``panda''}, \textit{``avast''}, \textit{``kaspersky''}, \textit{``eset''} or \textit{``norton''}), and social media used to contact the post creator (\textit{``skype''}, \textit{``discord''}, \textit{``email''}). From these results, a key takeaway is that users are mostly concerned with the product sold, its effectiveness and the type of crypter.
    
\begin{figure}[h!]
          \centering
          \includegraphics[width=\linewidth]{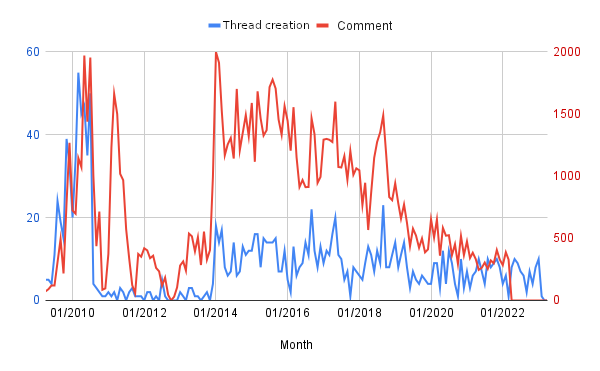}
          \caption{Thread creation and comments over time}
          \label{fig:byDate}
        \end{figure}

        Next, we analyze the activity on the forum and its evolution across time. We also attempt to correlate this activity with real-world events, such as cyberincidents or the irruption of the 2020 lockdown.
        Figure~\ref{fig:byDate} shows the activity month by month for both the comments left (in red) and threads initiated in the marketplace (in blue). As it can be observed the subforum gained rapid popularity from its creation, in 2009.

        In 2011 there are two significant events that might have led to the decrease in crypter activity, first the hacktivist group known as "LulzSec" leaked credentials and personal information of nearly 200,000 users from HackForums~\cite{pwnHF}. The forum reputation took such a hit that it has not really recovered ever since, not even close in terms of publications. This might correlate to the sudden drop by that year. Additionaly, as explored by Bhalerao et al.~\cite{bhalerao2018automatic}, in 2014 malware obfuscation alternatives became more frequent and crypters lost their popularity. Also, it's noteworthy a spike in late 2013, which we can not relate to any other public incident or event. Finally, in 2020 there is a slight increase in the activity, potentially derived from the lockdown effects~\cite{buil2021cybercrime}.
    
        The last approach investigated in the generic analysis is to get the number of posts asking for help or for a specific product instead of selling a crypter.
        A simple --yet effective-- way to achieve this classification is by searching for certain keywords usually used in the context of asking for help. Concretely, the keywords used are: ``help'', ``need'', ``advice'', ``advise'', ``buy'', ``request'', ``question'', ``looking for'', ``doubt'', ``seeking''.

        Out of 1 492 posts, 371 of them (24.87\%) are asking for something instead of selling a product. Examples of these threads are: ``NEED HELP/ASSISTANCE IN RE-FUDING A CVE2017-0199 EXPLOIT.'', ``PDF EXPLOIT NEEDED ASAP'', ``Looking for Private Crypt [Async RAT]'', ``Crypter NEEDED!! PLEASE READ!!'' or even ``I want to buy infected computers''. This shows that this community is a marketplace for offer and demand of these systems, confirming the role of crypters on the supply chain within the cybercrime ecosystem~\cite{bhalerao2019mapping}.

    \subsection{Top100 posts}
        
\begin{figure}[h!]
          \centering
          \includegraphics[width=0.5\linewidth]{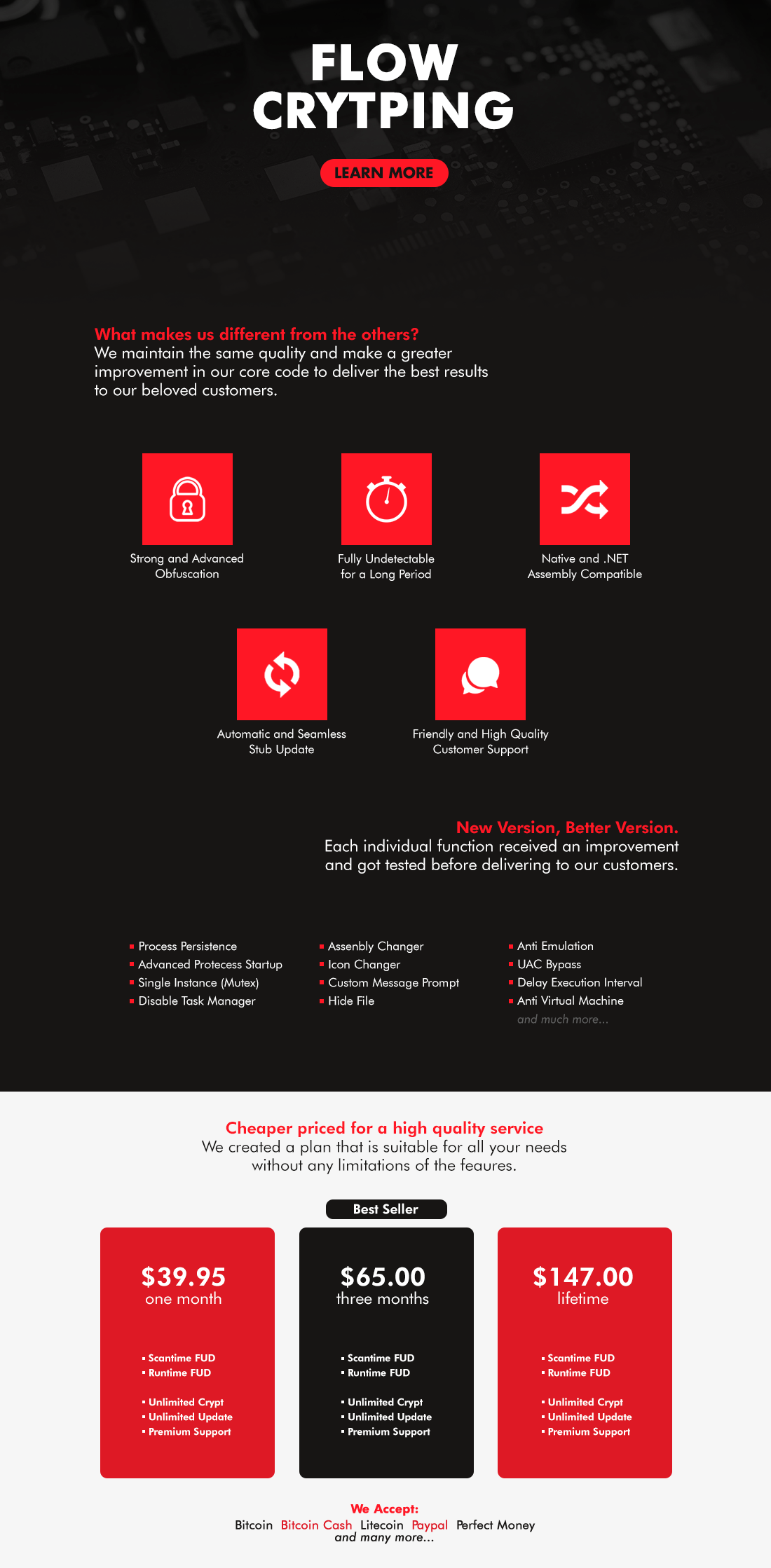}
          \caption{Example of leaflet used}
          \label{fig:folleto}
        \end{figure}

        To better understand the crypter-as-a-service ecosystem, we conduct a detailed, semiautomatic analysis of the top 100 posts (by number of views) to ensure we analyze the most significant and popular products of the subforum. Since there is no such thing as crypter specialization there is no other significant product outside our top 100 group. Since posts usually combine written text with image pamphlets, (Figure~\ref{fig:folleto} shows an example), this hardens the text extraction. We first attempt to use state-of-the art technologies for OCR (optical character recognition), i.e., \textit{tesseract}~\cite{tesseract}. However, the data extraction was inexact and led to multiple errors, hindering the automatic analysis of these images. Thus, we opted to conduct the analysis through manual inspection of these images.

        Even though these products are intended for malicious purposes, they are also commercial products being sold in a public marketplace. Thus, they include marketing techniques to attract customers, build a brand and gain some reputation. The marketing aspect is often provided in the pamphlet that summarize the specific features of the crypter, the prize and sales, accepted payment methods, or contact details. Also, they usually give away free vouchers after publishing a post and show the bypass rate to gain some trust in the community.
        Customer satisfaction is an important factor in most of the threads, they announce 24x7 quality customer support, reply to user comments in the forum and update the original post of the thread quite often to appear active and with a working and undetected stub. Despite all these efforts they are products with potential misuse being sold in a public website. From the creator's side, they often disengage from any responsibility by writing a disclaimer, typically forbidding any user from using their product for illegal activities.
        
        Out of the top 100 posts, 11 of them were currently closed or paused at the time of the analysis, and only 15 have been active since 2020, which is in accordance with the timeline analysis provided in the previous section.
        When it comes to the users creating those crypters, there are 13 users that have created two, three or four of them, the most remarkable of them is one user (User 2 in Table~\ref{tab:top100}), who is tied in the top one creators and three of his posts have been active in the past three years, even though two of them seem to be duplicates of the same crypter ``FLOW CRYPTER PRO V7''
        This user resembles the role of a reseller o ``commercial'', i.e., users who are only in charge of the marketing and selling of products, thus hiding the actual creator and allowing them to remain anonymous.

        We have previously commented on the fact that some crypters use leaflets to announce themselves as a well-marketed product. In the top 100, however, only 27 actually use this method, whereas the rest stick to a text-based description. This shows that the use of leaflets does not necessarily reflect a higher impact of the product.

        There are various similarities in terms of technical functionalities offered by the product (custom startup, customer icon, virtual machine and sandbox detection), platform and payment methods. The latter often rely on cryptocurrencies such as Bitcoin or Ethereum, and sometimes also PerfectMoney, which is an Internet payment system that doesn't force to verify its users allowing the peers to stay anonymized.

        In Table~\ref{tab:top100} we have compared the fifteen posts within the top100 that are still active since 2020, their overall position in the top 100, name, id of the creator (that has been anonymized), creation and last interaction years, price and the type of stub provided.
    
        \begin{table*}
            \caption{15 posts from the TOP 100 that remain active since 2020 (as of May'23)}
            \label{tab:top100}
            \resizebox{\linewidth}{!}{
                \begin{tabular}{cllcccclll}
                    \toprule
                        \multicolumn{1}{c}{Position}&\multicolumn{1}{c}{Name}&\multicolumn{1}{c}{Creator}&\multicolumn{1}{c}{Created}&\multicolumn{1}{c}{Last Comment}&\multicolumn{1}{c}{Views}&\multicolumn{1}{c}{Comments}&\multicolumn{1}{c}{Stub}&\multicolumn{1}{c}{Min cost}&\multicolumn{1}{c}{Max cost}\\
                    \midrule
                        3&ByteCrypter v3&User 1&2017&2023&205 721&3 282&Generator&35\$ 3months&60\$ lifetime\\
                        4&FLOW CRYPTER PRO V7&User 2&2017&2023&204 256&2 483&Standard&39,95\$ 1month&147\$ lifetime\\
                        11&CyberSeal&User 3&2017&2020&112 744&1 390&Standard&No info&No info\\
                        13&DATAPROTECTOR v4&User 4&2018&2020&109 893&1 865&Standard&50\$ 45days&300\$ lifetime\\
                        22&BetaCrypt&User 5&2014&2022&86 765&842&Private&210\$ 1month&No info\\
                        27&PURE CRYPTER&User 6&2021&2023&73 027&885&Standard&No info&No info\\
                        30&Data Encoder Crypter&User 6&2020&2023&66 704&909&Both&60\$ 1month Standard&175\$ 180days Private\\
                        32&RAZ PRIVATE CRYPTS&User 7&2020&2023&56 526&619&Private&25\$ 1crypt&285\$ 15crypts\\
                        46&Cassandra Crypter&User 8&2019&2021&43 196&533&Generator&24.99\$ 1month&No info\\
                        52&Code Protector&User 9&2019&2022&40 241&570&Standard&25\$ 1month&65\$ lifetime\\
                        61&STATIC CRYPT&User 10&2019&2022&27 663&359&Standard&50\$ 1month&500\$ lifetime\\
                        62&FLOW CRYPTER PRO V7&User 2&2018&2023&27 049&381&Private&20\$ 1crypt&399\$ 3months\\
                        63&PRIVATE CRYPTS&User 2&2020&2023&26 350&387&Private&20\$ 1crypt&399\$ 3months\\
                        65&TRILLIUM SECURITY FILE PROTECTOR V1.60&User 11&2020&2023&25 868&385&Both&125\$ 1month Standard&550\$ 1year\\
                        86&AtillaCrypt V2&User 12&2019&2020&16 119&280&Generator&35\$ 1month&120\$ 6months\\
                    \bottomrule
                \end{tabular}
            }
        \end{table*}

        As we can see there are three types of stubs: private, which allegedly is a completely different one for every user; standard, which is shared by all the users (or by every $N$ users); and the unique stub generator, that uses a standard generator with some parameters to output some kind of pseudo-private stub. A private stub implies more work than a standard one, which is reflected in the prize since it is always more expensive (sometimes even 10 times more). But the advantage it poses is that the less users are using the same stub, the lower the probability of it being fingerprinted and detected by AVs.

        It is very common for creators to provide new stubs each time they are detected so customers can re-crypt their malware in order to stay undetected as long as possible, they will usually notify them via the forum, on Telegram channels or other chosen methods.

    \subsection{Social Network}
        
        The latest analysis of this study is about the social connections established in the sub-forum dedicated to crypters on HackForums. We analyse the types of users based on their forum activity: how many post they create, and how many only comment on other posts, how many users interact with each post, and how often. This information is useful for multiple reasons. It allows to get a better knowledge on the most popular topics. It also informs whether there is some specialization among users, or whether users who sell crypters also participate in discussions on others, or they just stick to their own product.
    
        Table \ref{tab:users} shows some statistics of types of users present in the marketplace and how they interact with the forum. A vast majority of the users (98.23\%) only comment, instead of publishing threads. This is expected in a marketplace-like forum. Two-thirds of the users only interact with one post, which suggests the general tendency for low interaction with the subforum apart from buying and help-seeking. Lastly, we have the creators or users that publish one or more crypters, where the most common case is users that create one post and comment on several ones.

        \begin{table}[h!]
            \caption{Types of users in the marketplace}
            \label{tab:users}
            \resizebox{\columnwidth}{!}{
                \begin{tabular}{lcc}
                    \toprule
                        \multicolumn{1}{c}{Description}&\multicolumn{1}{c}{Number}&\multicolumn{1}{c}{Percentage}\\
                    \toprule
                        Only commenting one post&9 999&63.5\%\\
                        Commenting more than one post&5 467&34.72\%\\
                        \rowcolor{lightgray}Users that only comment&15 466&98.23\%\\
                        Creating and commenting one post&54&0.34\%\\
                        Creating multiple posts and commenting one&11&0.07\%\\
                        Creating one post and commenting multiple&154&0.98\%\\
                        Creating and commenting multiple posts&60&0.38\%\\
                        \rowcolor{lightgray}Users that create at least one post&279&1.77\%\\
                    \bottomrule
                        \multicolumn{1}{c}{Total}&15 745&100\%\\
                    \bottomrule
                \end{tabular}
            }
        \end{table}

\begin{figure}[ht]
          \centering
          \includegraphics[width=\linewidth]{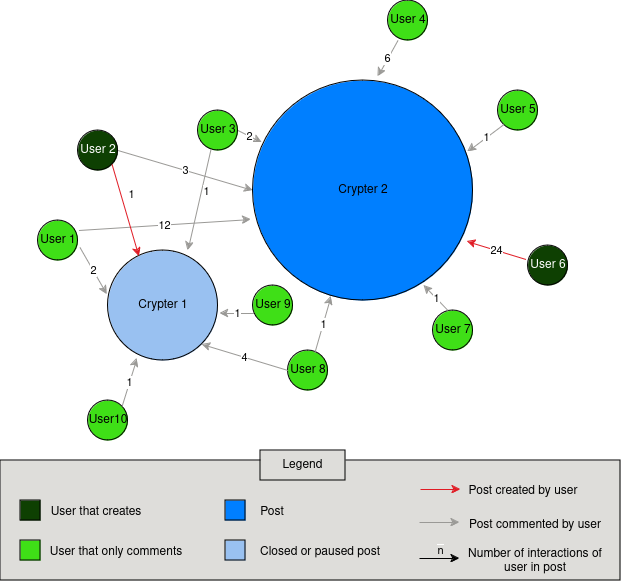}
          \caption{Screenshot of the social network graph}
          \label{fig:socialNetwork}
        \end{figure}

        To better reflect the interactions of users with posts, we have created an interactive graph of the social connections of posts and users within the top 100 posts. This allows to navigate and better analyze all these interactions in the marketplace in a visual way, focus on a single node and get the number of comments left on it, differentiate users that have created a post (and which is it) from those that have not, and the posts that have been closed from the ones that remain active. Figure~\ref{fig:socialNetwork} shows a representation of said graph, in it we can see how all that information mentioned is represented. This graph is fully interactive and available online~\cite{red}.
        This representation allows us to analyze particularities of the social network, e.g., by spotting groups of nodes that are separated, which would be related to specializations for some users, or by seeing the biggest post nodes we can find popular topics. We, however, observe that there are no anomalous nodes and they are all relatively similar. Also, the graph is homogeneous, which indicates that the community is equally distributed, with no particular niches.

        Additionally, we rely on metrics from graph theory, very important in the computer science field, such as centrality, to find the most important nodes in our social network with a given criteria~\cite{graphTheory}.

        The two chosen metrics are degree centrality, which in a directed graph like this one is defined as the combination of relations from a node (out-degree) and to a node (in-degree); and the eigenvector, which measures the relevance of a node from the importance of its neighbor nodes.

        The degree centrality in this particular scenario could be translated as the number of interactions to foreign posts (out-degree) from a user and the comments left in posts created by a user (in-degree) and by combining these we would get the most active and relevant users in the marketplace. The obtained ranking can be matched with the actual ``popularity'' property, or ``ranking'' of users in the forum so to speak, to figure out if there is any correlation between these two.

        Our preliminary study shows that in the same fashion as other analyses performed in this study, there is no actual relation between the in and out degree and the popularity of users, showing that the crypter marketplace is an isolated environment since the top ranked users by their degree are not the ones with most ``popularity'' within the forum, which might mean that they are not as active in the rest of marketplaces and discussions in the website. For example, user 13 has the highest ranking in the forum (6596) and it only has an out-degree of 3 with no in-degree whatsoever. Meanwhile, the user with the highest in-degree is user 4, with 4543 and only 1388 of popularity, almost a fifth of what user 13 has. When sorting by out-degree our top scorer is user 14 with 480 and an almost despicable, in comparison, popularity of 214. If we were, however to sort out users based on degree (combination of both in-degree and out-degree) user 4 would win again, meaning that creating popular posts such as ``DATAPROTECTOR v4'' (top 13) is the most significant factor when becoming a relevant user in the marketplace.
    
        The eigenvector on the other hand does give us more valuable results, as seen in Table~\ref{tab:eigenvector} the top 5 nodes with highest eigenvector values are present in the top posts that are still active since 2020, compared in Table~\ref{tab:top100}. They are not in order, or the top 5 posts in that list but that might be due to factors like the relevance of the users that comment on those posts or the number of different users interacting with the post rather than the raw number of views and comments of the post.

        \begin{table}[h!]
            \centering
            \caption{Eigenvector centrality}
            \label{tab:eigenvector}
            \resizebox{0.7\columnwidth}{!}{
                \begin{tabular}{clc}
                    \toprule
                        \multicolumn{1}{c}{Position}&\multicolumn{1}{c}{Name}&\multicolumn{1}{c}{Eigenvector}\\
                    \toprule
                        3&ByteCrypter v3&0.3964\\
                        4&FLOW CRYPTER PRO V7&0.3815\\
                        13&DATAPROTECTOR v4&0.2258\\
                        27&PURE CRYPTER&0.2183\\
                        11&CyberSeal&0.2012\\
                    \bottomrule
                \end{tabular}
            }
            \par
        \end{table}

        The main result of this social network analysis is the interactive graph seen in Figure~\ref{fig:socialNetwork}, but there are multiple ways to improve it, the first of them might be adding all the nodes. It is now restricted to the top 100 posts and users interacting with them due to computational limits, and also to ease visualization. But extending the graph with the entire dataset would provide a bigger picture. This format might benefit law enforcement units to look for any suspect in ongoing investigations and find created threads and interactions with other posts, but there is way more that can be done that wasn't completed for this research because it is not in our scope. One could improve the search bar to more effectively find and suggest nodes and add a side panel to view the interactions done with each post and the properties of the selected node, getting a very similar result to what BloodHound~\cite{bloodhound} facilitates in the Active Directory red team space.

\section{Case study: ByteCrypter}
\label{sec:technical}
    
    In an attempt to further explore this cybercrime landscape, we run one of the top sold products in the marketplace to corroborate that the offering features in the thread are in fact possible with the final crypter, and to compare the different binaries before and after obsuscating them through the crypter, so as to probe its success rate.

    We chose ByteCrypter v3~\cite{bytecrypter} as our sample because it is the top 3 post in the platform, the one with most eigenvector, as seen in Table~\ref{tab:eigenvector}, and has been active since 2020, so it is the best candidate to get a significant image on how a crypter operates. We got access to a released version from June 2022, already fingerprinted by the AV industry (the updated, most recent version, allegedly remains undetected from AVs).
    
    To check if the crypter actually works and evades any antivirus engine, we run the tool with both a malware and a regular, benign binary. For the malware, we consider a crypto-mining Monero wallet. Even if this `malware' could be used for benign purposes, it is commonly flagged as malicious by AVs~\cite{pastrana2019first}, and thus it serves well for our purpose. For the benign binary, we choose the standard Windows calculator application. By testing the crypter on both of them, we can study if the potential detection by AVs depends on the original binary, with its actual behaviour, or on the stub used incorporated by the crypter to obfuscate binary.

    \begin{figure}[h!]
      \centering
      \includegraphics[width=0.8\linewidth]{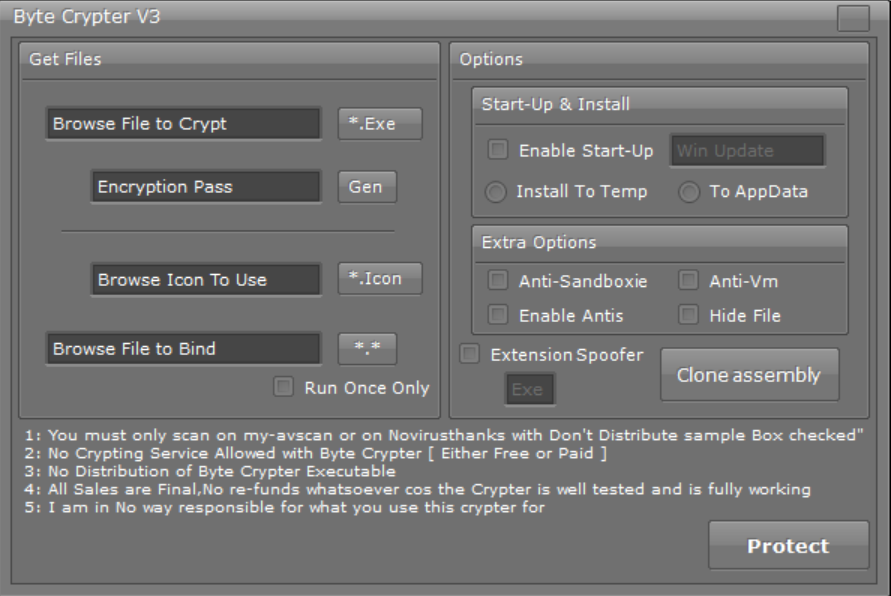}
      \caption{ByteCrypter v3 client GUI}
      \label{fig:bytecrypter}
    \end{figure}

    Upon opening the crypter, we can see what options are available from the Graphical User Interface (GUI) and they are in fact the same ones announced in the post: icon personalization, custom startup, sandbox and Virtual Machine (VM) detection and prevention along with others.
    
    We use the crypter to obfuscate both the Calculator and the cryptomining malware. Then, for the two of them, we uploaded both the original and obfuscated versions to \textit{VirusTotal}, and obtained interesting results. On the one hand, the original calculator binary was not detected by any of the antivirus engine, which is is not surprising at all because it is a very known binary signed by Microsoft. However, whereas it is not suspicious and non malicious, its crypted version was flagged as malicious by 54 of the 71 engines (76\%). This is most likely due to the fact that the version of the crypter tested dates back to June 2022, and the stub has been detected (and then fingerprinted by the AV industry) since then. Thus, it is a matter for future research obtaining an updated version of the crypter, allegedly FUD, to test whether it is detected or not. On the other hand, the cryptoming sample downloaded from the official website is detected as malicious by 22/68 security vendors out of the box (32\%). However, when crypted with ByteCrypter it is only marked by 36/66 (55\%), which is a surprising lower ratio of AVs than those detecting the obfuscated version of the Calculator. This shows that the application of the crypter actually modifies the original binary, but the resulting stub is detected differently by AVs.
        
    Overall, we observe that this version of ByteCrypter permits to bypass detection in only three antivirus engines. These are few cases comparing with the 14 engines that flagged malicious the crypted version of the cryptominer, and not the original, non encrypted version. This suggests that this outdated version does not allow for evading antiviruses, and thus strengthens the need for a dynamic, udpating service which is the current business model in most of the studied crypters.
    
    To conclude this case study, we observe that the studied crypter actually provides the features announced in the thread advert. However, without an official updated version of the stub, the resulting binaries, no matter what the actual originals are, are indeed detected as malicious by the antivirus engines.

\section{Related work}
\label{sec:related-work}
    
    \textbf{Cybercrime} is a relevant area of research, because both cybersecurity professionals and law enforcement officers need to understand the cybercriminal landscape and evolution, the attacks they use, how they share knowledge, and how to improve our defenses to stop the newest techniques~\cite{hughes2024art}. Cybersecurty has now established as a main national strategy~\cite{10nations}, and there is an increased  amount of response teams (or CERTs) or the incentives and cooperation with private agencies to improve the overall cyberdefense ecosystem. And as long as cyberdefenses are improved, cybercriminals also want to adapt in a never-ending cat-and-mouse game~\cite{hutchings2019displacing}. In the case of crypters, the need for updates became a necessity when anti viruses improved. 
    This way, malware operators can keep their businesses operative over time, for example for ransomware infections~\cite{ransom}. As such, the crypter-as-a-services is considered a Business-to-Business model and it fosters the entrepeneurship of criminal endeavours, since it allow to bypass one of the most complex barriers, i.e., the technical skill~\cite{bohme2021silicon}
    
    A important data source to study cybercrime are the \textbf{underground forums and markets}: they are public websites where new and/or experienced cybercriminals share knowledge, sell tools and form a community where they can gather into organized groups or even coordinate attacks. The seminal work by Motoyama et al.~\cite{undergroundForums} provided the first analysis on the social factor of these forums, how users interact in these communities and how it affects the forum itself. Indeed, previous studied have focused on HackForums, for example to understand new crime types such as eWhoring~\cite{eWhoring1,eWhoring2}, which involves the distribution of pornographic material in HackForums, or to study the evolution of key actors from their initial pursuits in the forum, to criminal activities~\cite{pastrana2018characterizing}.

    Other authors have particularly focused on malware and other \textbf{hacking tools} traded in these underground forums. The work by Valero and García~\cite{rats} reviews some of the most significant remote access trojan (known as RATs) along the years comparing its functionalities, the forums they used to spread and attacks where they were used.
    Regarding obfuscation, Efstratios et al.~\cite{avEvasion} analyzed multiple evasion techniques for antiviruses written in Go, Rust and C++ Also, they analyse the use of ChatGPT to explore its capabilities to generate malware. 
    Sembera et al. analysed one service providing obfuscation for malware in the Android ecosystem, studying the features and economy of such particular service ~\cite{vsembera2021cybercrime}. Finally, we refer to the survey by Muralidharan et al. for a description of the most current techniques for malware obfuscation for PE files~\cite{muralidharan2022file}.

    Despite previous efforts, mostly focused on the technical parts, we observe a research gap covering the operation and marketing of crypter-as-a-service ecosystem, and its social network. Our study contributes to enlightening this space by analyzing the marketplace in HackForums, and helps to shed light on this previously unexplored ecosystem.
        
\section{Discussions and conclusions}
\label{sec:conclusions}

    In this paper, we study the crypter-as-a-service ecosystem in HackForums. We observe that most of the the advertised products and services offer similar features in their adverts. Also, they offer similar properties, payment methods, customer support and close prices. This is, however, something expected, since technically a crypter is a standard concept, and leaves little gap for creativity and innovation. However, we also observe important differences in the popularity and attraction of the crypters. The differentiating part is, most probably, the way each crypter creates the stub (i.e., the encryption process), and also the customer support, i.e., the treatment given to customers once the service or product is delivered (e.g., quality of the support, how often they update stubs, or whether the actual stub is generic for all or a group of users or even private to each one of them). 
    Regarding targeted architectures, all the products we encountered are Windows and .NET based, probably due to Windows holding 28.08\% OS share for all devices\cite{osShare}, and 72.22\% of desktops\cite{desktopShare}. It is expected for malware developers to focus on the biggest amount of victims available, but this is not the only factor in place, Android, for example with 42.73\% OS share and 71.25\% in mobile is not a popular platform in crypter development because antiviruses are not as widely spread as it is in Windows and this is also the case for macOS, iOS and Linux distributions. There is no need for an antivirus evader if the most probable case is that there is none at all, malware developers and distributors rely on other techniques more suited to the platform such as hiding trojanized applications in the Play Store \cite{playStore} or unwanted preinstalled software~\cite{preInstalled}.
    
    In order to confirm this observation, future research need to extend this study over other forums and marketplaces, including those from the Dark Web, to discard a specialization only in HackForums. Also, it would be interesting to further explore the service provided, though this would involve additional economic, legal and ethical challenges for the service acquisition and analysis.

\section*{Acknowledgments}
    As part of the open-report model followed by the Workshop on Attackers \& CyberCrime Operations (WACCO), all the reviews for this paper are publicly available at \url{https://github.com/wacco-workshop/WACCO/tree/main/WACCO-2024}.
    This paper was partially supported by INCIBE grant  APAMCiber within the framework of the Recovery, Transformation and Resilience Plan funds, financed by the European Union-NextGenerationEU, and by grant TED2021-132170A-I00 from the Spanish Ministry of Science and Innovation, funded by MCIN/AEI/10.13039/501100011033, and the European Union-NextGenerationEU/PRTR.

\bibliographystyle{IEEEtran}
\bibliography{bibliography}

\begin{thebibliography}{10}
\providecommand{\url}[1]{#1}
\csname url@samestyle\endcsname
\providecommand{\newblock}{\relax}
\providecommand{\bibinfo}[2]{#2}
\providecommand{\BIBentrySTDinterwordspacing}{\spaceskip=0pt\relax}
\providecommand{\BIBentryALTinterwordstretchfactor}{4}
\providecommand{\BIBentryALTinterwordspacing}{\spaceskip=\fontdimen2\font plus
\BIBentryALTinterwordstretchfactor\fontdimen3\font minus \fontdimen4\font\relax}
\providecommand{\BIBforeignlanguage}[2]{{%
\expandafter\ifx\csname l@#1\endcsname\relax
\typeout{** WARNING: IEEEtran.bst: No hyphenation pattern has been}%
\typeout{** loaded for the language `#1'. Using the pattern for}%
\typeout{** the default language instead.}%
\else
\language=\csname l@#1\endcsname
\fi
#2}}
\providecommand{\BIBdecl}{\relax}
\BIBdecl

\bibitem{obfuscation}
I.~You and K.~Yim, ``Malware obfuscation techniques: A brief survey,'' 11 2010, pp. 297--300.

\bibitem{crimeBB}
\BIBentryALTinterwordspacing
S.~Pastrana, D.~R. Thomas, A.~Hutchings, and R.~Clayton, ``Crimebb: Enabling cybercrime research on underground forums at scale,'' in \emph{Proceedings of the 2018 World Wide Web Conference}, ser. WWW '18.\hskip 1em plus 0.5em minus 0.4em\relax Republic and Canton of Geneva, CHE: International World Wide Web Conferences Steering Committee, 2018, p. 1845–1854. [Online]. Available: \url{https://doi.org/10.1145/3178876.3186178}
\BIBentrySTDinterwordspacing

\bibitem{plugPlay}
\BIBentryALTinterwordspacing
R.~van Wegberg, S.~Tajalizadehkhoob, K.~Soska, U.~Akyazi, C.~H. Ganan, B.~Klievink, N.~Christin, and M.~van Eeten, ``Plug and prey? measuring the commoditization of cybercrime via online anonymous markets,'' in \emph{27th USENIX Security Symposium (USENIX Security 18)}.\hskip 1em plus 0.5em minus 0.4em\relax Baltimore, MD: USENIX Association, Aug. 2018, pp. 1009--1026. [Online]. Available: \url{https://www.usenix.org/conference/usenixsecurity18/presentation/van-wegberg}
\BIBentrySTDinterwordspacing

\bibitem{buil2021cybercrime}
D.~Buil-Gil, F.~Mir{\'o}-Llinares, A.~Moneva, S.~Kemp, and N.~D{\'\i}az-Casta{\~n}o, ``Cybercrime and shifts in opportunities during covid-19: a preliminary analysis in the uk,'' \emph{European Societies}, vol.~23, no. sup1, pp. S47--S59, 2021.

\bibitem{hughes2024art}
J.~Hughes, S.~Pastrana, A.~Hutchings, S.~Afroz, S.~Samtani, W.~Li, and E.~Santana~Marin, ``The art of cybercrime community research,'' \emph{ACM Computing Surveys}, vol.~56, no.~6, pp. 1--26, 2024.

\bibitem{avEvasion}
E.~Chatzoglou, G.~Karopoulos, G.~Kambourakis, and Z.~Tsiatsikas, ``Bypassing antivirus detection: old-school malware, new tricks,'' 2023.

\bibitem{vsembera2021cybercrime}
V.~{\v{S}}embera, M.~Paquet-Clouston, S.~Garcia, and M.~J. Erquiaga, ``Cybercrime specialization: An expos{\'e} of a malicious android obfuscation-as-a-service,'' in \emph{2021 IEEE European Symposium on Security and Privacy Workshops (EuroS\&PW)}.\hskip 1em plus 0.5em minus 0.4em\relax IEEE, 2021, pp. 213--226.

\bibitem{muralidharan2022file}
T.~Muralidharan, A.~Cohen, N.~Gerson, and N.~Nissim, ``File packing from the malware perspective: techniques, analysis approaches, and directions for enhancements,'' \emph{ACM Computing Surveys}, vol.~55, no.~5, pp. 1--45, 2022.

\bibitem{repo}
\BIBentryALTinterwordspacing
A.~de~la Cruz. (2023, Jun.) Hackforums crypters analysis repository. [Online]. Available: \url{https://github.com/PaquitoelChocolatero/HFCrypterAnalysis}
\BIBentrySTDinterwordspacing

\bibitem{bohme2021silicon}
R.~B{\"o}hme, R.~Clayton, and B.~Collier, ``Silicon den: Cybercrime is entrepreneurship,'' in \emph{Workshop on the Economics of Information Security (WEIS)}, 2021.

\bibitem{pastrana2019first}
S.~Pastrana and G.~Suarez-Tangil, ``A first look at the crypto-mining malware ecosystem: A decade of unrestricted wealth,'' in \emph{Proceedings of the Internet Measurement Conference}, 2019, pp. 73--86.

\bibitem{paquet2022motivations}
M.~Paquet-Clouston and S.~Garc{\'\i}a, ``On the motivations and challenges of affiliates involved in cybercrime,'' \emph{Trends in Organized Crime}, pp. 1--30, 2022.

\bibitem{caballero2011measuring}
J.~Caballero, C.~Grier, C.~Kreibich, and V.~Paxson, ``Measuring $\{$Pay-per-Install$\}$: The commoditization of malware distribution,'' in \emph{20th USENIX Security Symposium (USENIX Security 11)}, 2011.

\bibitem{crypters}
T.~Micro, ``Crypter - definition,'' https://www.trendmicro.com/vinfo/us/security/definition/crypter, 2013, online: Last accessed 07/28/23.

\bibitem{stubs}
\BIBentryALTinterwordspacing
D.~0x00sec. (2016, May) Crypters - instruments of the underground. [Online]. Available: \url{https://0x00sec.org/t/crypters-instruments-of-the-underground/386}
\BIBentrySTDinterwordspacing

\bibitem{MANKY20139}
\BIBentryALTinterwordspacing
D.~Manky, ``Cybercrime as a service: a very modern business,'' \emph{Computer Fraud \& Security}, vol. 2013, no.~6, pp. 9--13, 2013. [Online]. Available: \url{https://www.sciencedirect.com/science/article/pii/S1361372313700538}
\BIBentrySTDinterwordspacing

\bibitem{pastrana2018characterizing}
S.~Pastrana, A.~Hutchings, A.~Caines, and P.~Buttery, ``Characterizing eve: Analysing cybercrime actors in a large underground forum,'' in \emph{Research in Attacks, Intrusions, and Defenses: 21st International Symposium, RAID 2018, Heraklion, Crete, Greece, September 10-12, 2018, Proceedings 21}.\hskip 1em plus 0.5em minus 0.4em\relax Springer, 2018, pp. 207--227.

\bibitem{cyberCrime}
K.~Turk, S.~Pastrana, and B.~Collier, ``A tight scrape: methodological approaches to cybercrime research data collection in adversarial environments,'' in \emph{2020 IEEE European Symposium on Security and Privacy Workshops (EuroS\&PW)}, 2020, pp. 428--437.

\bibitem{pwnHF}
\BIBentryALTinterwordspacing
(2011, Jun.) {Have I been pwned? Pwned websites}. [Online]. Available: \url{https://web.archive.org/web/20151003211856/https://haveibeenpwned.com/PwnedWebsites}
\BIBentrySTDinterwordspacing

\bibitem{bhalerao2018automatic}
R.~Bhalerao, M.~Aliapoulios, I.~Shumailov, S.~Afroz, and D.~McCoy, ``Towards automatic discovery of cybercrime supply chains,'' 2018.

\bibitem{bhalerao2019mapping}
------, ``Mapping the underground: Supervised discovery of cybercrime supply chains,'' in \emph{2019 APWG Symposium on Electronic Crime Research (eCrime)}.\hskip 1em plus 0.5em minus 0.4em\relax IEEE, 2019, pp. 1--16.

\bibitem{tesseract}
\BIBentryALTinterwordspacing
(2023, Jun.) {tesseract}. [Online]. Available: \url{https://github.com/tesseract-ocr/tesseract}
\BIBentrySTDinterwordspacing

\bibitem{red}
\BIBentryALTinterwordspacing
A.~de~la Cruz. (2023, Jun.) Hackforums crypter social network. [Online]. Available: \url{https://paquitoelchocolatero.github.io/HFCrypterAnalysis/social-network/canvas.html}
\BIBentrySTDinterwordspacing

\bibitem{graphTheory}
\BIBentryALTinterwordspacing
A.~Majeed and I.~Rauf, ``Graph theory: A comprehensive survey about graph theory applications in computer science and social networks,'' \emph{Inventions}, vol.~5, no.~1, 2020. [Online]. Available: \url{https://www.mdpi.com/2411-5134/5/1/10}
\BIBentrySTDinterwordspacing

\bibitem{bloodhound}
\BIBentryALTinterwordspacing
A.~Robbins. The bloodhound gui. [Online]. Available: \url{https://bloodhound.readthedocs.io/en/latest/data-analysis/bloodhound-gui.html}
\BIBentrySTDinterwordspacing

\bibitem{bytecrypter}
\BIBentryALTinterwordspacing
(2022, Jun.) Hacking-software-v3 byte crypter v3.rar. [Online]. Available: \url{https://github.com/Zusyaku/Hacking-Software-V3/commits/main/Byte%20Crypter%20V3.rar}
\BIBentrySTDinterwordspacing

\bibitem{10nations}
A.~T. Odebade and E.~Benkhelifa, ``A comparative study of national cyber security strategies of ten nations,'' 2023.

\bibitem{hutchings2019displacing}
A.~Hutchings, S.~Pastrana, and R.~Clayton, ``Displacing big data: How criminals cheat the system,'' in \emph{The Human Factor of Cybercrime}.\hskip 1em plus 0.5em minus 0.4em\relax Routledge, 2019, pp. 408--424.

\bibitem{ransom}
I.~W. Gray, J.~Cable, B.~Brown, V.~Cuiujuclu, and D.~McCoy, ``Money over morals: A business analysis of conti ransomware,'' 2023.

\bibitem{undergroundForums}
\BIBentryALTinterwordspacing
M.~Motoyama, D.~McCoy, K.~Levchenko, S.~Savage, and G.~M. Voelker, ``An analysis of underground forums,'' in \emph{Proceedings of the 2011 ACM SIGCOMM Conference on Internet Measurement Conference}, ser. IMC '11.\hskip 1em plus 0.5em minus 0.4em\relax New York, NY, USA: Association for Computing Machinery, 2011, p. 71–80. [Online]. Available: \url{https://doi.org/10.1145/2068816.2068824}
\BIBentrySTDinterwordspacing

\bibitem{eWhoring1}
A.~Hutchings and S.~Pastrana, ``Understanding ewhoring,'' in \emph{2019 IEEE European Symposium on Security and Privacy (EuroS\&P)}.\hskip 1em plus 0.5em minus 0.4em\relax IEEE, 2019, pp. 201--214.

\bibitem{eWhoring2}
\BIBentryALTinterwordspacing
S.~Pastrana, A.~Hutchings, D.~Thomas, and J.~Tapiador, ``Measuring ewhoring,'' in \emph{Proceedings of the Internet Measurement Conference}, ser. IMC '19.\hskip 1em plus 0.5em minus 0.4em\relax New York, NY, USA: Association for Computing Machinery, 2019, p. 463–477. [Online]. Available: \url{https://doi.org/10.1145/3355369.3355597}
\BIBentrySTDinterwordspacing

\bibitem{rats}
V.~Valeros and S.~Garcia, ``Growth and commoditization of remote access trojans,'' in \emph{2020 IEEE European Symposium on Security and Privacy Workshops (EuroS\&PW)}, 2020, pp. 454--462.

\bibitem{osShare}
\BIBentryALTinterwordspacing
(2023, Jun.) {Operating System Market Share Worldwide {$\vert$} Statcounter Global Stats}. [Online]. Available: \url{https://gs.statcounter.com/os-market-share}
\BIBentrySTDinterwordspacing

\bibitem{desktopShare}
\BIBentryALTinterwordspacing
(2023, Jun.) {Desktop Operating System Market Share Worldwide {$\vert$} Statcounter Global Stats}. [Online]. Available: \url{https://gs.statcounter.com/os-market-share/desktop/worldwide}
\BIBentrySTDinterwordspacing

\bibitem{playStore}
S.~Ryu, ``{Goldoson: Privacy-invasive and Clicker Android Adware found in popular apps in South Korea {$\vert$} McAfee Blog},'' \url{https://www.mcafee.com/blogs/other-blogs/mcafee-labs/goldoson-privacy-invasive-and-clicker-android-adware-found-in-popular-apps-in-south-korea}, April 2023.

\bibitem{preInstalled}
\BIBentryALTinterwordspacing
J.~Gamba, M.~Rashed, A.~Razaghpanah, J.~Tapiador, and N.~Vallina{-}Rodriguez, ``An analysis of pre-installed android software,'' \emph{CoRR}, vol. abs/1905.02713, 2019. [Online]. Available: \url{http://arxiv.org/abs/1905.02713}
\BIBentrySTDinterwordspacing

\end{thebibliography}

\end{document}